\pdfoutput=1
\PassOptionsToPackage{unicode}{hyperref}
\PassOptionsToPackage{hyphens}{url}
\documentclass[
]{article}
\usepackage{lmodern}
\usepackage{amssymb,amsmath}
\usepackage{ifxetex,ifluatex}
\ifnum 0\ifxetex 1\fi\ifluatex 1\fi=0 
  \usepackage[T1]{fontenc}
  \usepackage[utf8]{inputenc}
  \usepackage{textcomp} 
\else 
  \usepackage{unicode-math}
  \defaultfontfeatures{Scale=MatchLowercase}
  \defaultfontfeatures[\rmfamily]{Ligatures=TeX,Scale=1}
\fi
\IfFileExists{upquote.sty}{\usepackage{upquote}}{}
\IfFileExists{microtype.sty}{
  \usepackage[]{microtype}
  \UseMicrotypeSet[protrusion]{basicmath} 
}{}
\makeatletter
\@ifundefined{KOMAClassName}{
  \IfFileExists{parskip.sty}{%
    \usepackage{parskip}
  }{
    \setlength{\parindent}{0pt}
    \setlength{\parskip}{6pt plus 2pt minus 1pt}}
}{
  \KOMAoptions{parskip=half}}
\makeatother
\usepackage{xcolor}
\IfFileExists{xurl.sty}{\usepackage{xurl}}{} 
\IfFileExists{bookmark.sty}{\usepackage{bookmark}}{\usepackage{hyperref}}
\hypersetup{hidelinks}
\usepackage[a4paper, margin=4cm]{geometry}
\usepackage{libertine,libertinust1math}
\usepackage{graphicx}
\usepackage{enumitem}

\usepackage[all]{hypcap}

\title{Designing a Visual Tool for Property Graph Schema Extraction and Refinement: An Expert Study
\\[1ex]
\large Capita Selecta Databases (2IMD05)}

\author{Nimo Beeren\\\href{mailto:n.beeren@student.tue.nl}{\texttt{n.beeren@student.tue.nl}}\\Eindhoven University of Technology\\Department of Mathematics and Computer Science}

\date{\today}

\begin{document}

\maketitle

\begin{abstract}
\noindent The design space of visual tools that aim to help people create schemas for property graphs is explored. Interviews are conducted with experts in the domain of property graphs and data management in general. Through this collaboration, we determine how a schema extraction tool can provide value. These insights are used to establish design requirements and design a UI prototype, which are then relayed back to the experts. Positive reactions were received, which encourage future work in the property graph schema space.
\end{abstract}

\section{Introduction}
A \textit{property graph} (PG) can be used to represent data as a directed, attributed multi-graph, in which nodes and edges carry a set of labels and a set of key-value pairs, known as properties~\cite{bonifati2018querying}. PGs are used as the underlying data model in many common graph database systems, such as Neo4j\footnote{\url{https://neo4j.com/}}, TigerGraph~\cite{deutsch2019tigergraph}, and Amazon Neptune~\cite{bebee2018amazon}.

By imposing a set of integrity constraints (\textit{schema}) on a database system, we ensure that there is some agreement between data consumers and producers regarding the type of data that is stored. The schema can be used to better understand the structure of data and how entities are connected, which is essential for data integration, analytics, and physical optimization of the database system. The Linked Data Benchmark Council (LDBC)\footnote{\url{https://ldbcouncil.org/}} concerns itself with standardization and the community around graph processing technologies. In particular, the Property Graph Schema Working Group (PGSWG)\footnote{\url{https://ldbcouncil.org/gql-community/pgswg/}} is currently working to standardize the notion of schemas for PGs.

Because of the flexibility of the PG model, graph databases are often used to store a large variety of data. Unfortunately, these heterogeneous collections of data do not always have a well-defined structure in the form of a schema, which can make them difficult to work with. Recently, work has been done to generate a schema for a given PG instance, in a process called \textit{schema extraction}~\cite{lbath2020schema, lei2021thesis}. Given the advantages of schemas and the practical lack of structure in many PGs, these techniques may provide significant value.

While schema extraction techniques may guarantee that the data matches the extracted schema, we cannot expect the schema to be optimal from a semantic perspective. Datamodelling is an inherently human task, which relies on a broad knowledge of concepts and their relations in the real world. Hence, we may want to refine the extracted schema to our liking. Furthermore, a schema may evolve over time as data sources and requirements change.

Besides schema extraction and refinement, we also want to facilitate exploration of an existing PG schema. This is essential when trying to understand a schema as a whole, and when deciding how to change it. To this end, we employ a graph visualization method which brings numerous cognitive benefits, such as enhancing pattern detection, ability to examine a large amount of information, and providing an abstract view of the data by omitting details~\cite{card1999information}.

\renewcommand*{\thefootnote}{\fnsymbol{footnote}}
To tackle this challenge, a team of three members was formed. To start, Bei Li\footnote[1]{Bei is currently associated with Google.} is interested in all work around schema extraction and commonly works with knowledge graphs. Next, we have George Fletcher\footnotemark[2], who is a researcher and professor interested in query languages and graph databases. Finally, Nimo Beeren\footnotemark[2] is a master student in Computer Science \& Engineering, with a focus on helping people build better software. 
\footnotetext[2]{George and Nimo are currently associated with Eindhoven University of Technology.}
Furthermore, the contribution of experts, whom are introduced in \autoref{sec:interviews}, was invaluable in the realization of this project.
\renewcommand*{\thefootnote}{\arabic{footnote}}

The contributions of this work are:
\begin{itemize}
    \item A written report of 7 interviews with 4 experts in the domain of property graph schema.
    \item A set of informal requirements for a tool designed to facilitate schema extraction and refinement, inspired by the interviews.
    \item A UI prototype following the design requirements.
\end{itemize}

The rest of this document is structured as follows. In \autoref{sec:interviews}, the key insights from interviews with domain experts are described. In \autoref{sec:requirements}, a prioritized list of requirements is established. In \autoref{sec:design}, the UI prototype is shown, its views are linked to requirements. In \autoref{sec:discussion}, we reflect together with the experts on the design choices. \autoref{sec:conclusion} concludes the paper.

\section{Interviews}
\label{sec:interviews}
To determine how a visual schema extraction tool could work in the real world, we conducted interviews with four experts from industry. Following a convenience sampling strategy, members of the PGSWG were approached and asked to participate. The interviews were unstructured and informal in nature, with the focus on generating hypotheses and exploring ideas. This section is based on the author's interpretation, and does not necessarily represent the opinions of the interviewee or the company they work for. A list of all interviews and the dates they took place is given in \autoref{sec:dates}.

\subsection{Dušan Živković}
After decades of work and numerous jobs with organisations in the finance industry, Dušan is currently an architect at Thought Machine\footnote{\url{https://thoughtmachine.net/}}. While his current engagement revolves around a relational database as a foundation, he uses graphs to model systems conceptually.

In Dušan's experience across the industry, the presence of a solid structure in data is anything but guaranteed. Many datasets are only partially documented, and the accuracy and detail of this documentation is often suboptimal. Furthermore, documentation that may once have been accurate, can fall victim to the effects of ad-hoc schema changes. As an example, Dušan explains the concept of \textit{semantic drift}, where the usage and semantics of a data attribute change over time, while documentation is not updated.

The lack of documentation, such as schemas, can be explained by the current culture in data management. In practice, the typical approach is application-first, rather than knowledge-first. Data is often wrangled into a shape that fits a particular purpose, at the cost of reusability. This gives rise to a critical pain point: having an information need, how does one produce a suitable query?

Focusing on graph databases, we discuss how they are different from relational databases. In Dušan's work, the ability to execute graph traversal queries is one of the main advantages of graph databases. The graph data model allows modeling causal relationships, which is not really possible in the relational model. This is particularly useful when modeling processes.

Even though data is often stored in relational databases, Dušan explains that he often combines relational data with graphs. In particular, he uses a graph to model dependencies between tables. These may be functional dependencies or ordering-dependencies, for example. In a functional dependency, one or more attributes functionally determine another attribute. In ordering-dependency, events depend on one another, sometimes as a simple one-after-another sequence, but more often in form of complex directed acyclic graphs.

When prompted, Dušan agrees that visualizing a graph schema may be useful. However, most visualizations don't scale; any sufficiently complex schema quickly becomes overwhelming when visualized all at once. In practice, most of the time people are interested in only a small part of the graph, or how to ``get from one place to another''. This idea is not limited to schema graphs, but could also apply to a data graph. The question is then what elements of the graph to show or hide? There is a trade-off between simplicity and completeness, where showing too much can make it too hard to understand the graph at a glance, and showing too little may require too much `drilling down'. To make the right compromise, a user study would be very valuable.

\subsection{Victor Lee}
When it comes to hands-on experience with property graph schemas, Victor is a prime candidate to interview. He works at TigerGraph\footnote{\url{https://tigergraph.com/}}, which is a company that takes a schema-first approach to property graph databases. This allows them to improve performance and scalability, as long as the data has a well-defined structure. Hence, TigerGraph customers could benefit from a tool that helps with discovering such a structure.

Victor mentioned that many people who are getting started with TigerGraph already have data in a relational database that they want to use. For this use case, TigerGraph offers a migration tool that allows mapping a relational schema to a property graph schema. This procedure creates a schema mapping from tables to nodes, and from columns to node properties. Afterwards, the data can be loaded directly into TigerGraph.

Considering schema extraction, a similar workflow could be imagined. The main difference is that the relational migration procedure depends on a predefined structure in the form of a relational schema, while the goal of schema extraction is to produce a schema given only data. However, neither of these two approaches can be expected to produce a perfect schema in all situations. Therefore, it may be desirable to refine the initially generated schema through human intervention.

Some common schema refinement tasks were identified, starting with basic operations such as adding and removing nodes and edges. Furthermore, the direction of edges may be changed, or we may choose to add reverse edges. Depending on the graph DB implementation, the direction of edges may determine the types of queries that are supported, e.g. it may not be possible to execute a traversal query from edge target to the source.

Furthermore, it may be useful to merge two types into one, especially when data is gathered from multiple sources, as there may be overlapping or redundant types. Conversely, a single type can be split into two. To illustrate the usefulness of splitting types, consider a \texttt{Person} node type with an optional property \texttt{parkingSpot}. From an ontological perspective, it may be desirable to split \texttt{Person} into an \texttt{Employee} with a required \texttt{parkingSpot}, and a \texttt{Guest} without a \texttt{parkingSpot}.

Other refinement tasks include escalating a property to a node type, which would make certain traversal queries easier. For example, we could find all pairs of entities associated with the same property value (when this property is escalated to a node). However, when many entities are associated with the same property value (as is often the case with nominal data types), a performance penalty may be incurred when escalating this property to a node.


\subsection{Juan Sequeda}
Juan works at data.world\footnote{\url{https://data.world/}}, where he concerns himself with data cataloging, knowledge graphs, and facilitating collaboration around data. One of the products he worked on is Grafo\footnote{\url{https://gra.fo/}}, a design tool for knowledge graphs.

To start the discussion, we talked about what makes schemas valuable in general. The answer depends largely on who you ask. A data engineer may be interested in the physical implementation of the schema, with the purpose of optimizing query efficiency. In contrast, a data scientist may be more interested in using the schema to express data analysis tasks in the form of queries. Furthermore, the role of \textit{knowledge scientist}~\cite{fletcher2020knowledge} was identified as someone who takes responsibility for the usability and reliability of data, by means of maintaining schemas, data cataloging, and creating data integration mappings~\cite{fletcher2009towards}. These three personas are summarized in \autoref{tab:personas}.

The sentiment that organizations are often too application-focused is also reflected by Juan. Instead, we should take a knowledge-first approach, prioritizing quality of data. In this sense, a schema is a primary form of knowledge. Related is the emerging discipline of \textit{DataOps}~\cite{ereth2018dataops}, which aims to improve quality of data and the processes around it. Again, the ultimate goal of these practices is to enable people to get the information they want from a database, as quickly as possible.

This is one of the goals Grafo aims to reach. Juan describes Grafo as an \textit{ontology editor}, which allows people to document the structure of their knowledge graph, as well as mapping data sources to entities in the graph. Grafo supports both RDF and property graph representations, allowing the user to add concepts (nodes), relationships (edges), attributes (properties), and specializations (subtype relationships). Relational data sources can be mapped to the knowledge graph, but this must be done manually by selecting tables or columns from the relational schema.

When discussing schema extraction, Juan was not immediately convinced of the value. He stressed that it should be verified that an extracted schema indeed helps people to find what they are looking for with less effort. An experiment could be conducted to test the time it takes people to come up with a correct query to answer a particular question. In any case, it is clear that a study on the practical value of schema extraction must include real users.

While schema extraction could be used to generate a schema where there previously was none, it could also be interesting to use schema extraction to refine an existing schema. This could be useful when the existing schema is incomplete or outdated. Of course, no schema is perfect, and determining which schema is better is a difficult and very human task. To aid in this task, Juan suggests building some sort of \textit{diff} between two schemas, showing in what ways these schemas are different.

\begin{table}[ht]
    \centering
    \begin{tabular}{l|p{0.6\linewidth}}
        \textbf{Persona} & \textbf{Description} \\
        \hline
        Data engineer & Interested in the physical implementation of the schema, optimizing query efficiency. \\
        Data scientist & Interested in using the schema to express data analysis tasks in the form of queries. \\
        Knowledge scientist & Takes responsibility for the usability and reliability of data, by means of maintaining schemas, data cataloging, and creating data integration mappings.
    \end{tabular}
    \caption{The three personas that were identified during the interview with Juan.}
    \label{tab:personas}
\end{table}

\subsection{Joshua Shinavier}
Joshua currently works at LinkedIn\footnote{\url{https://linkedin.com/}}, and previously lead the data standardization effort at Uber\footnote{\url{https://uber.com/}}. His expertise lies in the field of graph data management, and knowledge graphs.

We first discussed why the decision to standardize data was made at Uber. The main problem was that the datasets that were being used all had their own versions of common concepts, such as \textit{driver} or \textit{trip}. This made it difficult to maintain the company-wide knowledge graph, especially when introducing new data sources. Every time a new data source was added, it took considerable effort to map the original data schema to the knowledge graph schema. This became the bottleneck of the process, so they set out to unify the schemas of all data in use at the company.

As we also found in other interviews, the vast majority of datasets being used in knowledge graphs are not graph data. Commonly, data is stored in relational databases, or in a schemaless format such as JSON. It is notable that at Uber, nearly all datasets had some well-defined structure as well as globally unique entity identifiers. This was very valuable in the standardization process, but as we found, not all organizations are in the same situation.

We also touched on the topic of schema evolution. In Joshua's experience, it was very common for schemas to change over time. To keep track of schema history, the schema files were often checked in to a version control system such as Git. Before a schema change is published, a number of checks are performed to ensure the changes are backwards compatible. This is essential for them, because many services depend on the existing schema.

The rules for backwards compatibility limit what kinds of changes can be made to the graph schema. In general, we can only add things, not remove them. This means that many of the schema refinement functions that we identified earlier would violate the rules. However, we should consider the use case here. On the one hand, for an existing system running in production, backwards compatibility is often of importance. On the other hand, when creating an initial schema from scratch, or when the entire data stack is restructured, these rules may not be relevant.

It can be very useful to see how a schema evolved over time. This could be done by viewing a diff between two different versions of the schema. This diff could be visual, for example by showing a graph where nodes and edges are green when they are added, and red when they are removed. This approach may be useful for specific types of changes such as adding nodes or changing labels, but Joshua warns that not all types of changes are easy to represent visually. For example, displaying all properties that were changed may produce a view that is too cluttered to be useful.

As a more general solution to the schema diff problem, Joshua suggests creating a \textit{semantic diff}. Instead of displaying changes as visual graph elements, we describe the changes using natural language, for example `Added node \texttt{Employee}' or `Changed property type \texttt{Person.age} from \texttt{string} to \texttt{integer}'. Such a diff is both easy to parse by humans (like a visual diff), and scalable (like a line-by-line text diff).

\section{Design Requirements}
\label{sec:requirements}
Using the insights obtained from the interviews, we now establish a set of design requirements using the MoSCoW method~\cite{clegg1994case} for prioritization. Requirements are linked to the interview that inspired them, although not all requirements were explicitly discussed during the interviews. Furthermore, the requirements \ref{req:extract} and \ref{req:save} were established by the team, independently from the interviews.

\textit{Note: these requirements are further discussed in \autoref{sec:discussion}, which resulted in a change of priority for some requirements. For a summary of the changes, see \autoref{tab:requirement-changes}}.

\paragraph{Must have}
\begin{enumerate}[label=M\arabic*., ref=M\arabic*]
    \item\label{req:extract} Extract a PG schema given a PG database instance. --- Bei/George/Nimo
    \item\label{req:visualize} Visualize the schema graph. --- All
    \item\label{req:display-schema-text} Display the schema in textual form. --- Joshua
    \item\label{req:add-node} Add/remove a node to/from the schema graph. --- Victor/Juan
    \item\label{req:add-edge} Add/remove an edge to/from the schema graph. --- Victor/Juan
    \item\label{req:add-property} Add/remove a property to/from the schema graph. --- Victor/Juan
    \item\label{req:set-property-type} Set the data type of a property. --- Victor/Juan
    \item\label{req:save} Save the schema. --- Bei/George/Nimo
\end{enumerate}

\paragraph{Should have}
\begin{enumerate}[label=S\arabic*., ref=S\arabic*]
    \item\label{req:relation-visual-text} Indicate the relation between the visual and textual representation of the schema graph. --- Joshua
    \item\label{req:modify-schema-text} Modify a PG schema in textual form. --- Joshua
    \item\label{req:visual-diff} Display a visual diff between two PG schemas. --- Joshua/Juan
    \item\label{req:semantic-diff} Display a semantic diff between two PG schemas. --- Joshua/Juan
    \item\label{req:merge-nodes} Merge any pair of nodes. --- Victor
    \item\label{req:merge-edges} Merge any pair of edges. --- Victor
    \item\label{req:duplicate-node} Duplicate a node. --- Victor
    \item\label{req:duplicate-edge} Duplicate an edge. --- Victor
    \item\label{req:edge-direction} Set the direction of an edge. --- Victor
\end{enumerate}

\paragraph{Could have}
\begin{enumerate}[label=C\arabic*., ref=C\arabic*]
    \item\label{req:expand-nodes} Expand schema nodes one by one. --- Dušan
    \item\label{req:backwards-compatibility} Prevent backwards-incompatible schema changes when desired by the user. --- Joshua
    \item\label{req:edge-cardinality} Set the cardinality of edge types. --- Victor
    \item\label{req:escalate} Escalate a property to a node. --- Victor
    \item\label{req:subtype} Visualize sub/supertype relations. --- Victor
\end{enumerate}

\paragraph{Won't have}
\begin{enumerate}[label=W\arabic*., ref=W\arabic*]
    \item\label{req:mutate} Mutate data to conform to schema changes.
    \item\label{req:map-data-sources} Map data sources to the PG schema.
    \item\label{req:reverse-edge} Add an edge in the reverse direction for a specified directed edge in the schema. --- Victor
\end{enumerate}

\clearpage
\section{UI Design}
\label{sec:design}

\begin{figure}[ht]
    \centering
    \includegraphics[width=\textwidth]{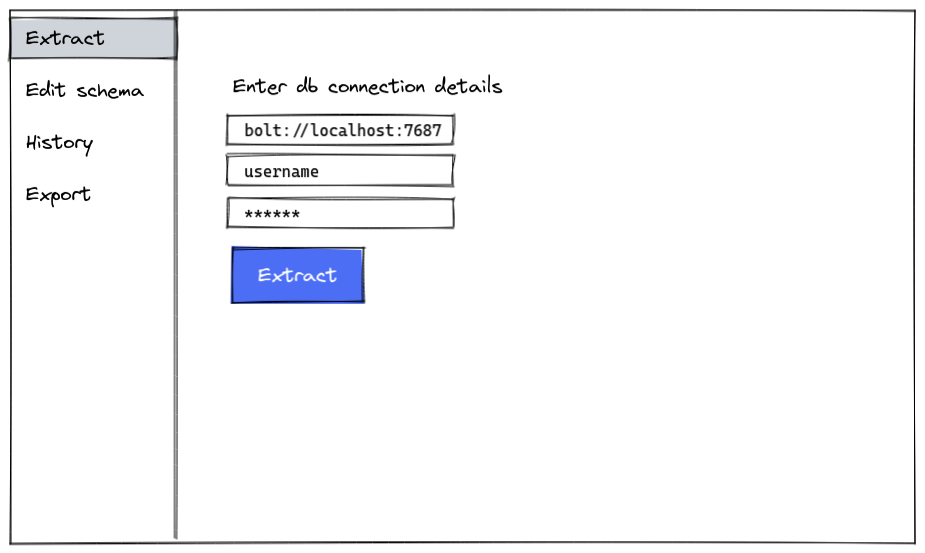}
    \caption{The \textit{Extract} screen allows connecting to a database instance and start schema extraction.}
    \label{fig:extract}
\end{figure}

\begin{figure}[ht]
    \centering
    \includegraphics[height=0.8\textheight]{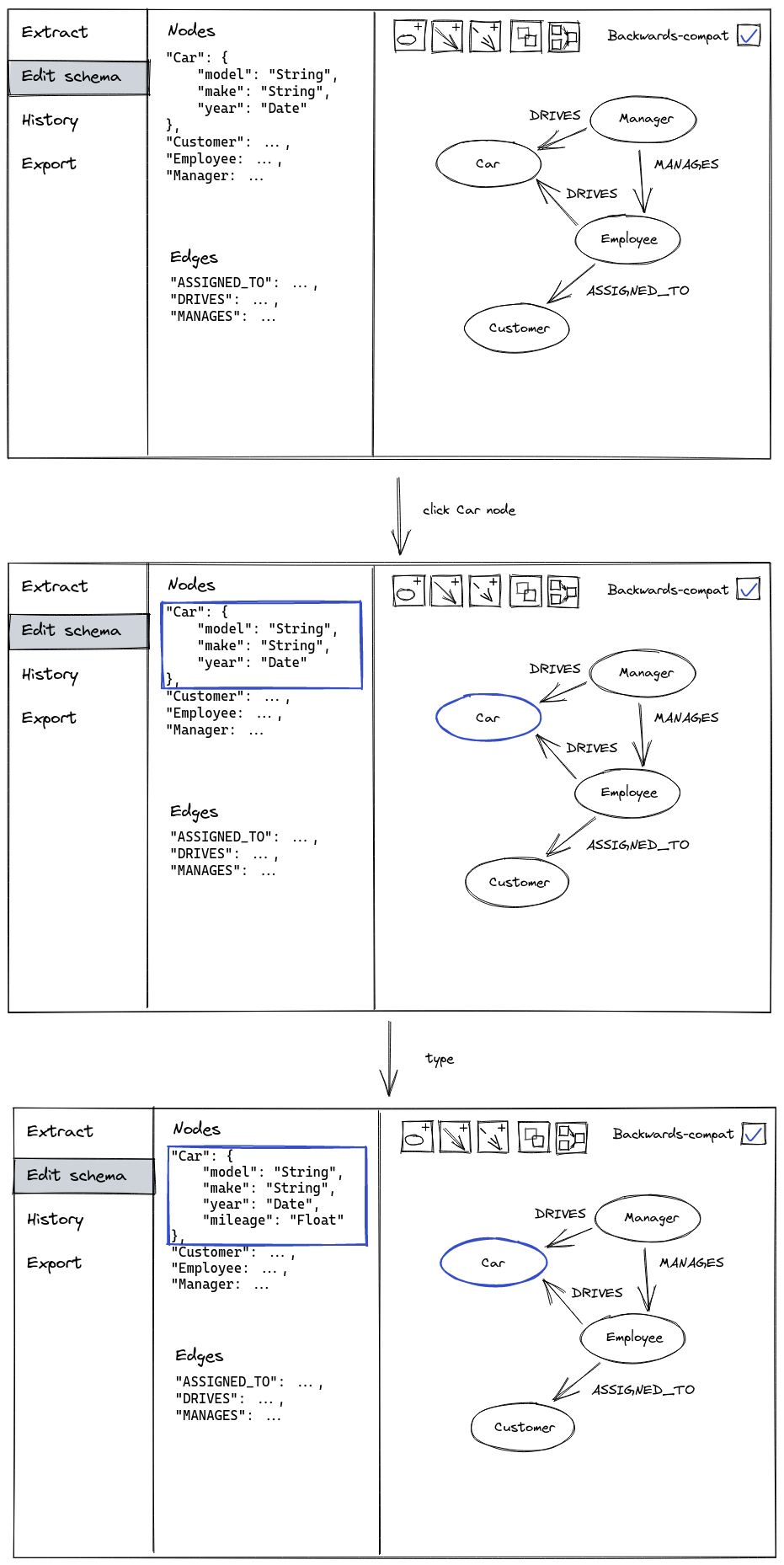}
    \caption{On the \textit{Edit schema} screen, the current schema can be refined. When a schema element is selected, the relationship between textual and visual representation of the schema is shown by a colored outline. The schema can be modified by typing in the text view, or using the buttons in the visual view. Backwards-incompatible changes can be prevented using a checkbox.}
    \label{fig:refine}
\end{figure}

\begin{figure}[ht]
    \centering
    \includegraphics[height=0.8\textheight]{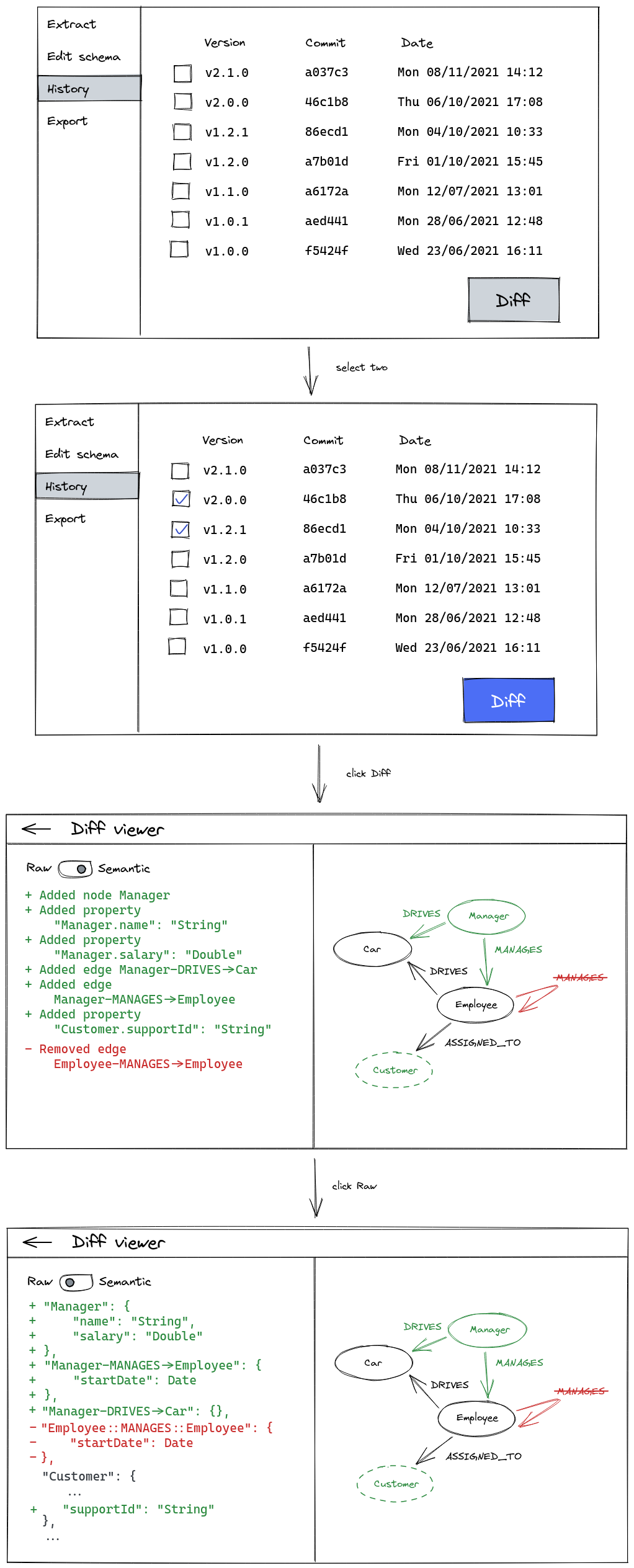}
    \caption{On the \textit{History} screen, two versions of the schema can be compared. After selecting the versions to compare, a diff view is shown. A switch allows toggling between the semantic and raw text view of the schema.}
    \label{fig:history}
\end{figure}

\begin{figure}[ht]
    \centering
    \includegraphics[width=\textwidth]{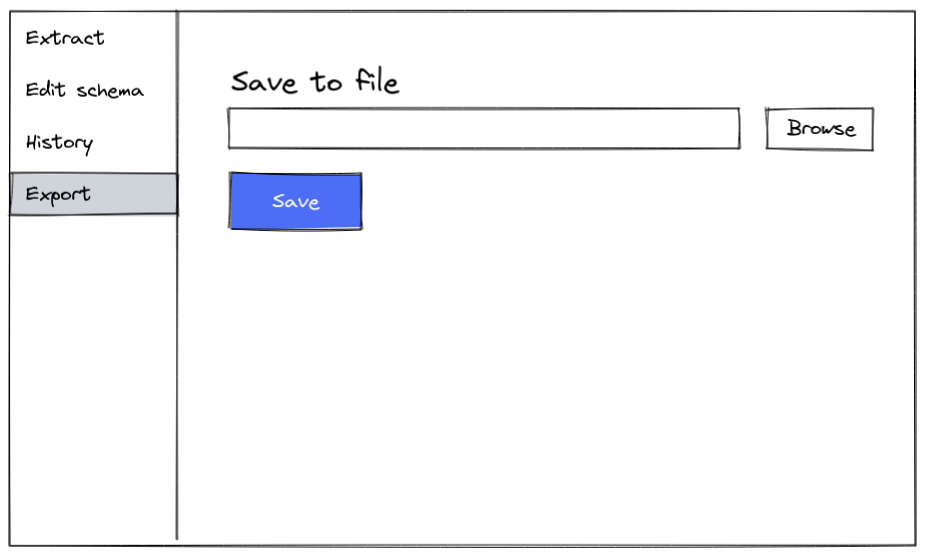}
    \caption{On the \textit{Export} screen, the current schema can be saved to a file.}
    \label{fig:export}
\end{figure}

Using the established design requirements as a guideline, a UI prototype was designed. The prototype consists of four screens: \hyperref[fig:extract]{\textit{Extract}}, \hyperref[fig:refine]{\textit{Edit schema}}, \hyperref[fig:history]{\textit{History}}, and \hyperref[fig:export]{\textit{Export}}. The relationships between screens and requirements are shown in \autoref{tab:screen-to-req}.

\begin{table}[ht]
    \centering
    \begin{tabular}{l|l}
        \textbf{Screen} & \textbf{Requirements} \\
        \hline
        \hyperref[fig:extract]{Extract} & \ref{req:extract}\\
        \hyperref[fig:refine]{Edit schema} & \ref{req:visualize}, \ref{req:display-schema-text}, \ref{req:add-node}, \ref{req:add-edge}, \ref{req:add-property}, \ref{req:set-property-type}, \ref{req:relation-visual-text}, \ref{req:modify-schema-text}, \ref{req:merge-nodes}, \ref{req:merge-edges}, \ref{req:duplicate-node}, \ref{req:duplicate-edge}, \ref{req:edge-direction},\\
        & \ref{req:backwards-compatibility}, \ref{req:subtype}\\
        \hyperref[fig:history]{History} & \ref{req:visualize}, \ref{req:display-schema-text}, \ref{req:visual-diff}, \ref{req:semantic-diff}, \ref{req:subtype}\\
        \hyperref[fig:export]{Export} & \ref{req:save} \\
        \textit{Unmet requirements} & \ref{req:expand-nodes}, \ref{req:edge-cardinality}, \ref{req:escalate}, \ref{req:mutate}, \ref{req:map-data-sources}, \ref{req:reverse-edge}\\
    \end{tabular}
    \caption{Mapping from screens in the UI prototype to the requirements they cover.}
    \label{tab:screen-to-req}
\end{table}

\newgeometry{margin=2cm}

\vspace*{\fill}
\vspace*{\fill}

\vspace*{\fill}
\vspace*{\fill}

\vspace*{\fill}
\vspace*{\fill}

\clearpage
\vspace*{\fill}
\vspace*{\fill}

\restoregeometry

\section{Discussion}
\label{sec:discussion}

After establishing the design requirements and designing the UI prototype, the experts were invited again to discuss the results. The setup of these interviews was similar to the first round. This time, we only talked with Dušan, Victor, and Juan.

\subsection{Schema refinement}
The feature that was most universally liked was the possibility to read and modify the schema in textual form (\autoref{req:display-schema-text}, \ref{req:modify-schema-text}). The main advantage of this approach is that it exposes the definitive \textit{source of truth} to the user. After all, the schema text is what is used by other applications, while the visual representation is only useful for humans. Another advantage is that it is easy to search through text by simply matching substrings (i.e. the well-known \texttt{Ctrl+F}). However, in this way we lose the semantics of the schema, making it impossible to discern node and property names, for example. Finally, the possibility to use common text manipulation methods such as copy/paste was appreciated.

Further discussion of the visual schema editor lead to some feedback on the \textit{merge} operation (\autoref{req:merge-nodes}, \ref{req:merge-edges}). It was not completely clear when this operation would be useful. Most experts indicated that they had not felt the need for this operation when they were doing schema design. However, when using a schema extraction algorithm, a situation could arise where a single entity was mistakenly identified as two different entities. But even in this context, the need for a merge operation seemed small enough to lower the requirement priority to \textit{Could~have}.

When explaining what the merge operation should actually do, it became clear there are actually multiple ways in which it could be interpreted. Indeed, the requirement does not precisely specify what the result of a merge should be. In type systems, we often distinguish between a type \textit{union} and \textit{intersection}. The original thought behind the merge operation was to do a union, i.e. combine properties from both types to create a type that allows all of those properties. As discussed, the usefulness of this operation seems limited. However, creating an intersection type may be useful in a different way. For example, we may wish to create a maximal supertype of a set of types, which is exactly what the intersection operation does. To summarize this point, the merge operation could be split into a union and intersection, though both would be relatively low-priority features.


\subsection{Visualization}
Regarding the visualization of the schema graph, feedback was mostly neutral because there are no really novel features. One feature that did spark a discussion was the possibility to \textit{expand schema nodes} (\autoref{req:expand-nodes}). First of all, it was not clear what this means. The original motivation for this requirement was to make it easier to navigate and explore large graphs. The idea was to avoid visual clutter by hiding most of the graph initially, and then allowing the user to select a node, after which its neighborhood would be shown.

A question that arose was about which schema elements to show initially. A possible solution would be to find a node or group of nodes that are somehow important, which could be achieved using a measure of centrality~\cite{borgatti2018analyzing}, for example. Another option would be to show everything by default, but let the user partition the schema into \textit{modules}, i.e. subgraphs which could be hidden or expanded at will (this is what Grafo does).

While this kind of user customization may be nice, it may only be useful in very specific cases. The experts agreed that in their experience, even the larger schemas only have around 100 nodes. While this is too large to make sense of when shown all at once, simple zooming and panning controls would be sufficient to explore the schema. Hence, we could consider demoting this requirement to \textit{Won't have}.

Finally, some improvement points regarding the user experience of the visual diff viewer were discussed. While the design uses green and red coloring to distinguish added and removed elements, this may not work for people who are not able to distinguish these colors well. A simple solution would be to add symbols like \texttt{+} or \texttt{-} to the visual diff, similar to the text diff. This would improve accessibility and usability for many people.

\subsection{Future work}

We now have a list of design requirements and a UI prototype based on them. We should keep in mind that the experts we interviewed were sampled from a single group (PGSWG), and this sample is not necessarily representative of all people who may benefit from a schema extraction tool. In future work, more interviews could be conducted to reach other target groups, such as application developers, data scientists, and anyone else who is concerned with schema design.

Furthermore, it is important to realize that the hypotheses presented in this work have not been validated outside the expert interviews. To increase confidence in the validity of the results, a user study could be done. This way, the impact of the tool could be quantified, by measuring e.g. the time it takes a user to produce a satisfactory schema for a given property graph instance.

\subsection{Summary}

As a result of this discussion, the priority of some requirements was changed. These changes are summarized in \autoref{tab:requirement-changes}.

\begin{table}[ht]
    \centering
    \begin{tabular}{l|l|l}
        \textbf{Requirements} & \textbf{Old priority} & \textbf{New priority} \\
        \hline
        \ref{req:merge-nodes}, \ref{req:merge-edges} & Should have & Could have \\
        \ref{req:expand-nodes} & Could have & Won't have
    \end{tabular}
    \caption{Requirements which had their priority changed as a result of the discussion.}
    \label{tab:requirement-changes}
\end{table}

In addition, the following opportunities for improvement were found:

\begin{itemize}
    \item The \textit{merge} operation could be split into two operations: \textit{union} and \textit{intersection}.
    \item The visual diff viewer could be made more accessible.
\end{itemize}

\section{Conclusion}
\label{sec:conclusion}
We investigated if and how schema extraction methods for property graphs could be applied to provide value. By talking to experts, we found that one of the main problems when working with data lies in producing a query for a particular question about data, though a schema can make this much easier. Given that graph data often lacks a complete schema, we see schema extraction as a potentially valuable tool.

We created design requirements and a UI prototype, which received encouraging reactions from experts. Most of the features were not directly related to schema extraction, but instead focused on refining an existing (extracted) schema. We found that textual and visual representations of the schema can be combined to enable easy exploration as well as manipulation. Furthermore, we identified some common operations on schema graphs. Finally, we found that a simple graph visualization was deemed sufficient to explore the schema in most cases.

Our results may stimulate future work on the design, implementation, and analysis of schema extraction tools. The world of graph data management is evolving quickly, and we believe that more tools around graph schema are necessary to meet the needs of data engineers, data scientists, and knowledge scientists everywhere.

\clearpage
\begin{appendix}
    \section{Interview Dates}
    \label{sec:dates}
    \begin{table}[ht]
        \centering
        \begin{tabular}{l|l}
            \textbf{Interviewee} & \textbf{Date}\\
            \hline
            Dušan Živković & 30-09-2021\\
            Victor Lee & 04-10-2021\\
            Juan Sequeda & 06-10-2021\\
            Joshua Shinavier & 15-10-2021\\
            Victor Lee & 12-11-2021\\
            Juan Sequeda & 12-11-2021\\
            Dušan Živković & 12-11-2021
        \end{tabular}
        \caption{The list of interviews with the date they took place.}
        \label{tab:dates}
    \end{table}
\end{appendix}

\clearpage
\bibliographystyle{plain}
\bibliography{main}

\end{document}